\documentclass{article}
\usepackage{iclr2021_conference}
\usepackage{times}
\usepackage{graphicx}
\usepackage{amssymb}
\usepackage{url}
\usepackage{color}
\usepackage{amsmath}
\usepackage{stmaryrd}
\usepackage{tikz}
\usepackage{listings}
\usepackage{amsthm}
\usepackage{hyperref}
\usepackage{url}
\usepackage{booktabs}
\usepackage{multirow}
\usepackage{siunitx}
\usepackage{makecell}
\usepackage{wrapfig}
\usepackage{authblk}

\newtheorem{definition}{Definition}

\newcommand{\ten}[1]{\mathcal{#1}}
\newcommand{\mat}[1]{\mathbf{#1}}

\iclrfinalcopy

\vspace{-25pt}
\title{On-FPGA Training with Ultra Memory Reduction: A Low-Precision Tensor Method} 
\vspace{-50pt}
\author{Kaiqi Zhang\textsuperscript{1}, Cole Hawkins\textsuperscript{2}, Xiyuan Zhang\textsuperscript{3}, Cong Hao\textsuperscript{4}, Zheng Zhang\textsuperscript{1}\\
\vspace{-5pt}
\textsuperscript{1} Department of ECE, UC Santa Barbara. E-mail: \texttt{\{kzhang70, zzhang01\}@ucsb.edu}\\
\textsuperscript{2}Department of Mathematics, UC Santa Barbara. E-mail: \texttt{colehawkins@ucsb.edu}\\
\textsuperscript{3}Department of ECE, Carnegie Mellon University. E-mail: \texttt{xiyuanzh@andrew.cmu.edu} \\
\textsuperscript{4}School of ECE, Georgia Institute of Technology. E-mail: \texttt{callie.hao@gatech.edu} 
}

\date{}

\begin{document}

\maketitle
\begin{abstract}
Various hardware accelerators have been developed for energy-efficient and real-time inference of neural networks on edge devices. 
However, most training is done on high-performance GPUs or servers, and the huge memory and computing costs prevent training neural networks on edge devices. 
This paper proposes a novel tensor-based training framework, which offers orders-of-magnitude memory reduction in the training process. 
We propose a novel rank-adaptive tensorized neural network model, and design a hardware-friendly low-precision algorithm to train this model. 
We present an FPGA accelerator to demonstrate the benefits of this training method on edge devices. Our preliminary FPGA implementation achieves $59\times$ speedup and $123\times$ energy reduction compared to embedded CPU, and $292\times$ memory reduction over a standard full-size training.
\end{abstract}

\section{Introduction} 
Modern neural networks consume huge memory and computing resources during both training and inference. This challenge is compounded by increasing demands for on-device training to preserve data privacy~\citep{teerapittayanon2017distributed} and to increase energy efficiency. Energy-efficient and low-latency inference with limited hardware resources have been well studied at both the algorithm~\citep{lecun1990optimal,neklyudov2017structured,lebedev2014speeding,hinton2015distilling,han2015deep,zhou2017incremental} and hardware levels~\citep{chen2016eyeriss, chen2019t, zhang2015optimizing, hao2019fpga, xu2020autodnnchip}. 
However, training on resource-constrained hardware remains an open challenge. 
Most on-device training algorithms utilize low-precision optimization~\citep{courbariaux2015binaryconnect,hubara2017quantized} to reduce the computing and memory cost per parameter in the training process. However, the cost reduction is limited to {\it a single order of magnitude} even if the most recent ultra low-precision 4-bit training~\citep{sun2020ultra} is employed. 

{\bf Paper Contributions.} This paper presents, for the first time, an {\it end-to-end} neural network training framework on FPGA with {\it orders-of-magnitude memory reduction}. This is achieved by developing a low-precision tensorized training framework. 
We propose a rank-adaptive tensorized training model, which employs a Bayesian method for automatic tensor rank determination and model compression in the training process. 
We further reduce the memory by employing low-precision computation to train the tensorized model. We use our algorithm to train a two-layer neural network on a Xilinx MPSoC, which stores all model parameters on chip, achieves $59\times$ speedup and $123\times$ energy reduction compared to embedded CPU, and $292\times$ memory reduction compared to original model.
\section{Preliminary}

\vspace{-5pt}
We use Tensor-Train Matrix (TTM) decomposition \citep{oseledets2011tensor} to compactly represent the neural network parameters in the training process. 

\begin{definition}
\label{def: tensor train matrix}
Let $\mat{W}\in\mathbb{R}^{J\times I}$ be the weight matrix in a fully connected layer and let $I=\prod_{n=1}^d I_n,J=\prod_{n=1}^d J_n$ be a factorization of its dimensions. We reshape $\mat{W}$ into a tensor $\ten{W}$ with dimensions $J_1\times\dots\times J_d\times I_1\times\dots\times I_d$~\citep{novikov2015tensorizing}.
The tensor-train matrix (TTM) factorization expresses each data element of $\ten {W}$ as a series of matrix products. 
\[
\ten{W} \left( j_1,\dots,j_d,i_1,\dots,i_d \right) = \ten{G}_1(:j_1,i_1,:)\ten{G}_2(:,j_2,i_2,:)\cdots \ten{G}_d (:,j_d,i_d,:)
\]
\end{definition}
Each TT-core factor $\ten{G}_n\in \mathbb{R}^{R_{n-1}\times J_n \times I_n \times R_{n} }$ is an order $4$-tensor, and $R_0$$=$$R_d$$=1$. The vector $\mat{R}=(R_0,R_1,\cdots,R_d)$ is called the TT-rank. This TTM format requires $\sum_{n=1}^d R_{n-1}J_n I_nR_{n}$ parameters to represent $\mat{W}$, which is much smaller than the original number of variables $\prod_n J_nI_n$. We may expect huge memory reduction if the TTM format can be employed in training, but it is challenging to decide the TT rank $\mat{R}$ in practice. Existing methods often use a fixed-rank training~\citep{novikov2015tensorizing,calvi2019compression,khrulkov2019tensorized} which require combinatoral rank searches and multiple training runs. Therefore, they are not suitable for one-shot on-device training.
\section{Low-Precision Rank-Adaptive Tensorized Neural Network}
To enable memory-efficient, one-shot and on-device training, this section proposes a rank-adaptive tensorized training method with low-precision optimization. 
 
\subsection{Rank-Adaptive Tenosirzed Training}

To simplify the descriptions, we assume that the parameters in each layer are represented with $d$ TT core factors $\boldsymbol \theta =\{ \ten{G}_n \in \mathbb{R}^{R_{n-1} \times J_n \times I_n \times R_n}\}_{n=1}^d$. Here $R_n$ is an initial rank for dimension $n$, and it is larger than the actual rank parameter that will be determined later in training. Our method applies to general cases that have multiple layers with each layer being parameterized by some TT cores.  We propose to train a tensorized neural network with the following rank-shrinkage model:
\begin{align}
\footnotesize
\min {\cal L}\left(  \boldsymbol \theta, \{ \boldsymbol{\lambda}_n\}_{n=1}^{d-1} \; |{\cal D}\right)  & = \frac{1}{|{\cal D}|}\sum_{\{\mat{x}_i,y_i\} \in {\cal D}} \mathrm{CE}\left( f\left(x_i, \boldsymbol \theta  \right), y_i \right) +g\left( \boldsymbol \theta, \{ \boldsymbol{\lambda}_n\}_{n=1}^{d-1} \right) \\
\text{with}\; g(\cdot) & = \sum \limits_{n=1}^{d-1} \sum \limits_{r_n=1}^{R_n} \left[ \frac{\|\ten{G}_n(:,:,:, r_n)\|_F^2}{\boldsymbol \lambda _n (r_n)} + \frac{1 + R_{n-1}I_nJ_n}{2} \log\left( \boldsymbol \lambda_n (r_n)\right ) \right]
\normalsize
\label{eq:reg1}
\end{align}
where ${\cal D}$ is the training data set, $\mathrm{CE}(\cdot, \cdot)$ is a cross-entropy loss, and $\boldsymbol \lambda_n \in \mathbb{R}^{R_n}$ denotes a group of rank-controlling hyper-parameters for TT core $\ten{G}_n$. 

\paragraph{Rank Reduction.} Our objective function is the negative log-posterior of the Bayesian model in \cite{hawkins2020towards}. 
Empirical results shows this prior leads to good trade-off between model size and accuracy, and it is hyper-parameter-free which eliminates the need of hyper-parameter fine-tuning.
Every element of subtensor $\ten{G}_n(:,:,:, r_n)$ (which is obtained by fixing the $4$-th index of $\ten{G}_n$) is equipped with a zero-mean Gaussian prior with a variance $\boldsymbol \lambda_n (r_n)$. Each element of $\boldsymbol \lambda_n$ is further assumed to have a Log-Uniform prior distribution, such that some elements of $\boldsymbol \lambda_n$ will approach $0$ with a high probability. In practice, once $\boldsymbol \lambda_n(r_n)$ becomes very small the variables in subtensor $\ten{G}_n(:,:,:, r_n)$ shrink to $0$, and they can be removed from $\ten{G}_n$, leading to a rank reduction.

\subsection{Training Quantized Tensor Factors}


To further reduce the memory and computing cost, we use BinaryConnect~\citep{courbariaux2015binaryconnect} to perform low-precision training. BinaryConnect keeps the real values of all low-precision parameters in a buffer. In each iteration, the gradients are accumulated in the buffer, and the low-precision parameters are updated by quantizing the buffer. Let a  real-valued TT factor be $\ten G_n$, denote its quantized variant as $\tilde {\ten G}_n$ and $Q(\cdot)$ as the quantization function, the update process for each TT core is
\begin{align}
    {\ten G}_n^{(t+1)}  = {\ten G}_n^{(t)} - \eta_t \nabla_{\tilde{\ten G}_n^t}{\cal L}\left( \{ \tilde{\ten G}_n^t\}_{n=1}^d, \{ \boldsymbol{\lambda}_n\}_{n=1}^{d-1} \; |{\cal B}\right), \ \quad
    \tilde{\ten G}_n^{(t+1)} =Q( {\ten G}_n^{(t+1)} ),
    \label{eq:bc2}
\end{align}
where ${\cal B} \in {\cal D}$ is a batch of randomly sampled training data, and $t$ is an index of the low-precision stochastic gradient-descent iteration. In iteration $t$, the hyper-parameters are updated as:
\begin{equation}
   \boldsymbol \lambda_n^t (r_n) = \frac{2}{1+R_{n-1}I_nJ_n} \|\ten{G}_n^t(:,:,:, r_n) \|_F^2.
    \label{eq:lamb}
\end{equation}


The quantization function is not differentiable.
Therefore, we use the straight-through estimator (STE) \citep{bengio2013estimating} to approximate the gradient of a quantization function. Specifically we use the gradient of a smooth function instead of the original non-differentiable activation in the backpropagation. 
This work uses a clipped ReLU as the STE. 
For ReLU activation function, denote $\mathbf{1}(\cdot)$ as the indicator function, the backpropagation rule can be written as
$\frac{\partial}{\partial x}=\mathbf{1}(x \geq 0 )\frac{\partial}{\partial y}.$
We use 4 bits to represent TT factors, 8 bits for activations and bias, and 16 bits for the gradients.
Since the processing elements (PE) are shared between forward and backward propagations, they are designed to handle 16-bit activations or gradients and 4-bit TT factors. During forward propagation only the 8 least significant bits (LSB) of the 16 bits are used. Therefore our trained model can be deployed for inference on edge devices with only 8 bits for activation and bias as well as 4 bits for TT factors.


\subsection{Automatic Scale Selection}

In fixed-point representations, each variable needs to be scaled carefully to avoid overflow or large quantization errors.
Furthermore, a bit shift is needed if the scaling factors differ by $2^i \times$.
In the training process, the values of activation functions and gradients can vary by several orders of magnitude. To handle this issue, we determine the scaling factor on the fly. The scaling factors of all variables are enforced to be $2^k$. We allow different scaling factors for each activation, gradient or intermediate result but share scaling factors across different data samples and different neurons of the same layer.
The scaling factors of the TT factors are fixed. To determine the scaling factor of the activation and gradients, we track the mean of their absolute values during training, and we enforce it to be in the range $[0.1, 0.3]$ by dynamically adjusting the scaling factor. This allows a small margin to avoid overflows, while making the most use of the hardware bits to reduce quantization errors.

\section{FPGA Implementation}


\begin{figure}[t]
    \centering
    \resizebox{0.99\columnwidth}{!}
{
    \begin{tikzpicture}
    \draw [draw=black, fill=blue!10] (0, 5) rectangle (8, 5.8);
    \node at (4, 5.4) {DRAM (training samples, activation and gradients)};
    \draw [draw=black, fill=green!10] (0, 0.6) rectangle (8, 4.8);
    \draw [draw=black, fill=blue!10] (0.2, 0.8) rectangle (6, 1.8);
    \node at (3.1, 1.3) {BRAM (Model parameters)};
    \begin{scope}[xshift=0]
        \draw [draw=black, fill=red!40] (0.2, 2) rectangle (3, 3);
        \node at (1.6, 2.7) {PE1};
        \node at (1.6, 2.3) {\small(forward\&backward)};
        \draw [draw=none, fill=blue!30] (0.2, 3.2) -- (0.2, 3.8) -- (1.7, 3.8) -- (1.5, 3.2) -- cycle;
        \draw [draw=none, fill=orange!40] (3, 3.2) -- (3, 3.8) -- (1.7, 3.8) -- (1.5, 3.2) -- cycle;
        \draw [draw=black] (0.2, 3.2) rectangle (3, 3.8);
        \node at (1.6, 3.5) {Ping-pong buffer};
        \draw [draw=black, fill=yellow!60] (0.2, 4) rectangle (3, 4.6);
        \node at (1.6, 4.3) {Load \& store};
        
        \draw [->] (1.6, 1.8) -- (1.6, 2);
        \draw [->] (2.2, 3) -- (2.2, 3.2);
        \draw [<-] (1, 3) -- (1, 3.2);
        \draw [->] (2.2, 3.8) -- (2.2, 4);
        \draw [<-] (1, 3.8) -- (1, 4);
        \draw [->] (2.2, 4.6) -- (2.2, 5);
        \draw [<-] (1, 4.6) -- (1, 5);
    \end{scope}
    \begin{scope}[xshift=3cm]
        \draw [draw=black, fill=red!40] (0.2, 2) rectangle (3, 3);
        \node at (1.6, 2.7) {PE2};
        \node at (1.6, 2.3) {\small(forward\&backward)};
        \draw [draw=none, fill=blue!30] (0.2, 3.2) -- (0.2, 3.8) -- (1.7, 3.8) -- (1.5, 3.2) -- cycle;
        \draw [draw=none, fill=orange!40] (3, 3.2) -- (3, 3.8) -- (1.7, 3.8) -- (1.5, 3.2) -- cycle;
        \draw [draw=black] (0.2, 3.2) rectangle (3, 3.8);
        \node at (1.6, 3.5) {Ping-pong buffer};
        \draw [draw=black, fill=yellow!60] (0.2, 4) rectangle (3, 4.6);
        \node at (1.6, 4.3) {Load \& store};
        
        \draw [->] (1.6, 1.8) -- (1.6, 2);
        \draw [->] (2.2, 3) -- (2.2, 3.2);
        \draw [<-] (1, 3) -- (1, 3.2);
        \draw [->] (2.2, 3.8) -- (2.2, 4);
        \draw [<-] (1, 3.8) -- (1, 4);
        \draw [->] (2.2, 4.6) -- (2.2, 5);
        \draw [<-] (1, 4.6) -- (1, 5);
    \end{scope}
    \draw [draw=black, fill=red!40] (6.2, 3.2) rectangle (7.8, 4.6);
    \node at (7, 4.1) {PE3};
    \node at (7, 3.7) {\small(backward)};
    \draw [->] (6.5, 4.6) -- (6.5, 5);
    \draw [<-] (7.5, 4.6) -- (7.5, 5);
    \draw [draw=black, fill=green!20] (6.2, 0.8) rectangle (7.8, 1.8);
    \node at (7, 1.5) {ARM};
    \node at (7, 1.1) {Cortex-M};
    \node at (7, 2.5) {MPSOC};
    \draw [dashed] (8.3, 0.7) -- (8.3, 5.7);
    \begin{scope}[xshift=8.5cm, yshift=3.1cm]
        \draw [draw=black] (0.8, 2.6) rectangle (3.8, 2.2);
        \node at (2.3, 2.4) {$\ten Z_k$};
        \draw [draw=black] (0, 0.6) rectangle (0.5, 2);
        \node at (0.25, 1.3) {$\ten G_l$};
        \foreach \x in {0, 2}
        \foreach \y in {0.6, 1.3}{
            \draw [->] (\x+2, \y+0.3) -- (\x+2, \y); 
            \draw (\x+1.5, \y+0.3) rectangle (\x+2.5, \y+0.7); 
            \node at (\x+2, \y+0.5) {\small MACC};
            \draw [->] (\x+\y/3+0.8, 2.2) -- (\x+\y/3+0.8, \y+0.6) -- (\x+1.5, \y+0.6);
        }
        \foreach \y in {0.6, 1.3}{
            \draw [->] (0.5, \y+0.5) -- (1.5, \y+0.5); 
            \draw [->] (2.5, \y+0.5) -- (3.5, \y+0.5); 
        }
        \draw [draw=black] (1.5, 0.2) rectangle (4.5, 0.6);
        \node at (3, 0.4) {$\ten Z_{k+1}$};
    \end{scope}
    \draw [dashed] (8.3, 3.2) -- (13.2, 3.2);
    \begin{scope}[xshift=8.5cm, yshift=0.5cm]
        \draw [draw=black] (1, 2.6) rectangle (4, 2.2);
        \node at (2.5, 2.4) {$\ten Z_k$};
        \draw [draw=black] (0, 0.6) rectangle (0.5, 2);
        \node at (0.25, 1.3) {$\ten G_l$};
        \foreach \x in {0, 2}
        \foreach \y in {0.5, 1.2}{
            \draw [->] (\x+1.2, \y+1) -- (\x+1.2, \y+0.7); 
            \draw (\x+0.8, \y+0.3) rectangle (\x+1.8, \y+0.7); 
            \node at (\x+1.3, \y+0.5) {\small MACC};
            \draw [->] (\x+1.8, \y+0.4) -- (\x+1.9+\y/3, \y+0.4) -- (\x+1.9+\y/3, 0.6);
        }
        \foreach \y in {0.5, 1.2}{
            \draw [->] (0.5, \y+0.5) -- (0.8, \y+0.5); 
            \draw [->] (1.8, \y+0.5) -- (2.8, \y+0.5); 
        }
        \draw [draw=black] (1.5, 0.2) rectangle (4.5, 0.6);
        \node at (3, 0.4) {$\ten Z_{k+1}$};
    \end{scope}
    \end{tikzpicture}
        }
    \caption{Left: overall view of the FPGA accelerator. Right: the structures of PE1 (top) and PE2 (bottom). MACC means ``multiply and accumulator". }
    \label{fig:overall}
\end{figure}
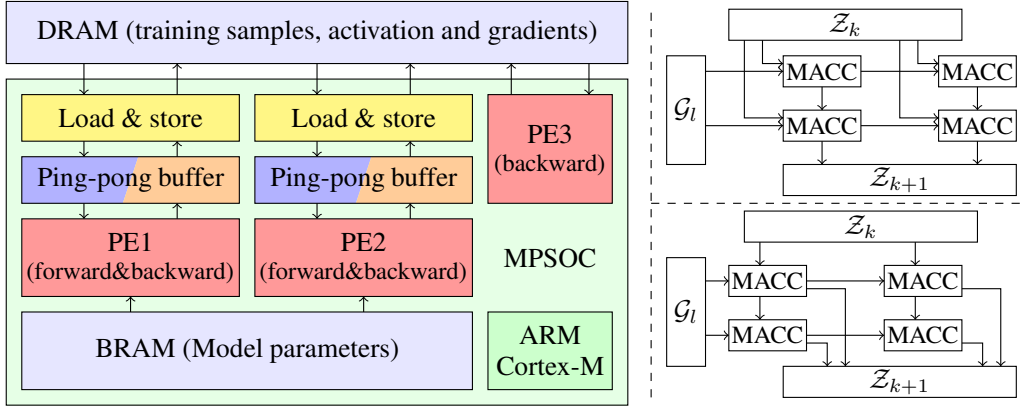

This section describes the FPGA implementation for our on-device tensorized training method. The overall FPGA design is shown in Fig~\ref{fig:overall}. The detailed computations of forward and backward propagation are given in the Appendix.
During training, the data samples, activations, and gradients are stored in the off-chip DRAM. Due to our low-rank tensorization, all the model parameters may be stored in the on-chip BRAM.
The overall training involves three steps: forward propagation, backward propagation, and model parameter update.
The forward and backward propagations run on the FPGA programmable logic;
the TT factors $\ten G_n$'s and rank parameters $\boldsymbol\lambda_n$'s are updated on the embedded ARM core, which usually take less than 1\% of the total computing time. 
We design three processing elements (PEs) for the forward and backward propagation:
PE1 and PE2 are shared by the forward and backward propagation and can take advantage of different data locality during tensor contraction. PE3 is dedicated to computing the outer products in a backward propagation.

PE1 and PE2 are shown in Fig.~\ref{fig:overall}. The computation involves three steps: loading a tensor slice from DRAM to the on-chip buffer, performing the multiply-and-accumulate (MAC) operations, and sending the results back to DRAM. With ping-pong buffers, the (most time-consuming) second step can be executed in parallel with other steps. We require only two key computational kernels to perform forward propagation. We describe the kernels of the first two PEs, and provide the details of the forward and backward pass in the Appendix. PE1 is used for a two-index tensor contraction which contains the last dimension of both operands. By reshaping the $d$-dimensional tensors into 3-dimensional tensors without permutation, the computation can be written element-wise as 
\begin{equation}
    \ten Z_{k+1}(a,d) = \sum_{b,c} \ten Z_k(a,b,c)\ten G_{l}(b,d,c).
    \label{eq:pe1 contraction}
\end{equation}
The second PE (PE2) performs a tensor contraction along a single dimension that is not the last:
\begin{equation}
    \ten Z_{k+1}(a,d,c) = \sum_{b} \ten Z_k(a,b,c)\ten G_{l}(b,d).
    \label{eq:pe2 contraction}
\end{equation}
Here $\ten{Z}_{k+1}$ and $\ten{Z}_k$ denote the intermediate results, and $\ten{G}_l$ is a reshaped TTM factor. Both PEs have 128 MACs that operate in parallel. Only order-$2$ and order-$3$ tensor contractions are required due to a reshape operation (see Table~\ref{tab:pe} in the Appendix). In a forward propagation, the mode index $l=d-k$; in a backward propagation, $l=k+1$. In PE1, we split the first operand based on the first dimension of $\ten{Z}_k$. We parallelize the computing along the last dimension (associated with index $c$) by a factor of 16, and along the first dimension (w.r.t. index $a$) by a factor of 8. This allows each element in the first operand to be shared across 8 multipliers.
To simplify the design, we enforce the dimension size of index $c$
to be a multiplier of 16, which is equivalent to enforcing the last dimension of both input and output tensors to be a multiplier of 16.
Similarly, in PE2, the first operand is split based on dimensions associated with indices $a$ and $c$. The computations associated with indices $c$ and $d$ are parallelized by a factor of 16 and 8, respectively. 
We introduce PE3 to perform outer products. The throughput of this step is bounded by the memory bandwidth of storing, so we parallelize the computing along the last dimension $I_d$ 
only by a factor of 16. The elements of the second operand is cached, while the first operand is read from DRAM directly. A total of 16 multipliers calculate the product of elements from the two operands and the results are written to DRAM directly without caching, because they are not further used in this step.



\section{Experiments and results}

We implement our low-precision rank-adaptive tensorized training on an Avnet Ultra96-V2 FPGA board and use it to train a two-layer neural network to classify the FashionMNIST dataset. We implemented the neural network in HLS codes, yet they can be easily modified to fit different network structures. The detailed experimental setup, including the structure of neural network and the hyperparameters, is provided in the Appendix \ref{sec:exp}.
As shown in Table \ref{tab:fm}, our method achieves $294\times$ memory reduction for the model parameters compared with the standard non-tensorized training.

\begin{table}[t]
    \centering
    \vspace{-1cm}
    \caption{Fashion MNIST training result.}
    \begin{tabular}{c|ccccc}
    \hline
    Method & \thead{Training\\ accuracy} & \thead{Testing\\ accuracy} & \thead{Model \\parameters} & \thead{Memory in bits} & \thead{Memory\\reduction}\\
    \hline
    Vanilla & 95.75\% & 89.27\% & $4.67\times 10^5$ & $1.49\times 10^7$ & N/A\\
    \hline
    Floating, w/o prior& 92.54\% & 88.03\% & $1.48\times 10^4$ & $4.74\times 10^5$ & 31.4$\times$\\
    Fixed, w/o prior & 88.31\% & 86.67\% & $1.48\times 10^4$ & $6.13\times 10^4$ & 243$\times$\\
    \hline
    Floating, w/ prior & 90.17\% & 87.88\% & $1.08\times 10^4$ & $3.46\times 10^5$ & $43.1\times$\\
    Fixed, w/ prior (proposed) & 85.45\% & 84.86\% & $1.22 \times 10^4$ & $5.11\times 10^4$ & 292$\times$\\
    \hline
    \end{tabular}
    \label{tab:fm}
\end{table}

The hardware resource utilization is listed in Table~2. 
We compare the time and memory usage of tensorized neural network training on FPGA and on an embedded processor, a Raspberry Pi 3B with Quad Core 1.2GHz ARM processor. We use the Pytorch and Tensorly modules to implement our training algorithm on the embedded processor. For the FPGA we set the clock rate to 100MHz. The forward and backward propagation takes 0.09s per batch of 64 samples, where the embedded processor takes 5.34s. The estimated power consumption of our design is 1.2W compared to 2.5W of Raspberry Pi, indicating that our FPGA accelerator is $123\times$ more energy-efficient. 
\begin{table}[t]
\begin{center}
\caption{Resource utilization of two-layer tensorized neural network.}
\begin{tabular}{c|ccccc}
    \hline
    resource & LUT & FF & DSP & BRAM & power\\
    \hline
    used        & 56131 & 30155 & 278 & 77 & 1.2W\\
    available   & 70560 & 141120 & 360 & 432 & -\\
    utilization & 79.55\% & 21.37\% & 77.22\% & 17.82\% & -\\
    \hline
\end{tabular}
\end{center}
\label{tab:resources}
\end{table}

\section{Conclusion and Future Work}
We have proposed a low-precision tensor method to train neural networks on edge devices. By end-to-end compressed training, our approach produce an ultra-compact model from scratch while saving significant hardware resources during the training. Our algorithm uses a rank-adaptive approach to determine model complexity, and it has achieved $292\times$ memory reduction compared to the baseline model with only a $4.5\%$ testing accuracy loss on Fashion MNIST. The FPGA implementation has achieved $59\times$ speedup and $123\times$ energy reduction than the training on an embedded CPU.

In the future we plan to: (1) further optimize the low-precision tensorized training algorithm and its FPGA implementation, and (2) demonstrate on-device training for large-size neural networks.

\bibliographystyle{iclr2021_conference}
\bibliography{ref,cole_simods}

\newpage
\appendix

\section{Detail about tensor contraction}
\subsection{Forward propagation}

To perform the forward propagation we take a vector $\mat{x}$ and the TTM cores $\{\ten{G}_n\}_{n=1}^d$ of the tensorized weight matrix $\mat{W}$. The forward propagation operation is 
\begin{equation}
    \label{eq: forward prop}
    \mat{y} = \mat{W}\mat{x}
\end{equation}
We detail the tensorized forward and backward passes where $\ten{X}$ is achieved by reshaping $\mat{x}$ and $\ten{Y}$ is the tensorized output.

The forward and backward propagation of neural network involves tensor contraction. Denote the input dimensions as $I_1, I_2, \dots, I_d$, output dimensions as $J_1, J_2, \dots, J_d$, and the rank as $R_1, R_2, \dots R_{d-1}$. the forward propagation involves the following computation: 
\begin{align}
    \ten Z_1(i_1,i_2,\dots,r_{d-1},j_d) 
    & =\sum_{i_d}\ten X(i_1,i_2,\dots i_d) \times \ten G_d(r_{d-1},j_d, i_d) 
    \label{eq:ct1}\\
    \ten Z_2(i_1, \dots, r_{d-2},j_{d-1},j_{d}) 
    &= \sum_{i_{d-1}r_{d-1}}
    \ten Z_1(i_1,i_2,\dots,r_{d-1},j_d) \times \ten G_{d-1}(j_{d-1},r_{d-1},r_{d-2},i_{d-1})
    \label{eq:ct2}\\
    &\dots \nonumber\\
    \ten Y(j_1,j_2,\dots j_d)
    &= \sum_{i_1}\ten Z_{d-1}(i_1,r_1,j_2,\dots, j_d) \times \ten G_1(j_1,r_1,i_1)  \label{eq:ct3}
\end{align}    
In these expressions, we denote $\ten X$ as the input to this layer, $\ten G_n$ as TTM core factors, $\ten Y$ as the output and $\ten Z_i$ as an intermediate result.

\subsection{Backward propagation}
In back propagation, there are two tasks:
\begin{itemize}
    \item To compute the gradients with respect to the inputs of the layer.
    \item To compute the gradients with respect to the compressed model parameters (TTM core factors) of the layer.
\end{itemize}

To compute the gradients with respect to the inputs, we denote $\hat{\ten X}$ as the gradient of input to this layer and $\hat{\ten Y}$ as the gradient of the output of this layer. The computation is shown below: 
\begin{align}
    \ten Z_1(i_1,r_1,j_2,\dots,j_d)
    &= \sum_{j_1}\hat{\ten Y}(j_1,j_2,\dots,j_d) \times \ten G_1(j_1,i_1,r_1) 
    \label{eq:ct4}\\
    \ten Z_2(i_1,i_2,r_2,\dots,j_d)
    &= \sum_{r_1j_2}\ten Z_1(i_1,r_1,j_2,\dots,j_d) \times \ten G_2(r_1,j_2,i_2,r_2) 
    \label{eq:ct5}\\
    &\dots\nonumber\\
    \hat{\ten X}(i_1,i_2,\dots,i_d) 
    &= \sum_{j_d}\ten Z_{d-1}(i_1,\dots,i_{d-1},r_{d-1},j_d) \times \ten G_d(r_{d-1},j_d,i_d) \label{eq:ct6}
\end{align}
The first equation is to compute the gradients directly. The second is to compute the gradients with respect to full weights and then accumulate them and compute the gradients with respect to the TTM factors. The former method is more efficient if the batch size is small and the compressed model is small, while the latter is more efficient otherwise. In our work, we start training with a large rank initial guess (i.e., a larger model), so the latter method is more efficient. The first step, computing the gradient w.r.t. the full weight matrix (denoted as $\hat{\ten W}$), requires a simple outer product which is computed by PE3:
$$
\hat{\ten W}(j_1,i_1,j_2,i_2,\dots,j_d,i_d)=
\ten X(i_1,i_2,\dots,i_d) \times \hat{\ten Y}(j_1,j_2,\dots,j_d) 
$$
The first operand of this PE is the input to this layer during forward propagation, and the second operand is the gradient of the output. After the gradient has been accumulated in a batch, the gradient with respect to the factors denoted by $\hat {\ten G}_i$ can be computed by contracting the gradient of full weight with the tensor factors:
\begin{align}
&\ten Z_{1,1}(j_1,i_1,j_2,i_2,\dots,j_{d-1},i_{d-1},r_{d-1}) =
\sum_{j_d, i_d}
\hat{\ten W}(j_1,i_1,j_2,i_2,\dots,j_d,i_d) \times \ten G_d(r_{d-1},j_d,i_d) \label{eq:ct8}\\
&\ten Z_{2,1}(j_1,i_1,j_2,i_2,\dots,j_{d-2},i_{d-2},r_{d-2})\nonumber\\
&\hspace{1cm}=
\sum_{j_{d-1}, i_{d-1}, r_{d-1}}
\ten Z_{1,1}(j_1,i_1,j_2,i_2,\dots,j_{d-1},i_{d-1},r_{d-1}) \times \ten G_{d-1,1}(r_{d-2},j_{d-1},i_{d-1}, r_{d-1}) \label{eq:ct13}\\
&\dots\nonumber\\
&\hat{\ten G_1}(j_1,i_1,r_1) 
=\sum_{j_2,i_2,r_2}\ten Z_{d-2,1}(j_1,i_1,j_2,i_2,r_2) \times \ten G_2(r_1,j_2,i_2,r_2) \label{eq:ct9}\\
&\hat{\ten G}_2(r_1,j_2,i_2,r_2) =
\sum_{j_1, i_1} \ten Z_{d-2,1}(j_1,i_1,j_2,i_2,r_2) \times \ten G_1(j_1,i_1,r_1)
&  \label{eq:ct10}\\
&\ten Z_{d-3,2}(r_1, j_2, i_2, j_3, i_3, r_3) = \sum_{j_1, i_1} \ten Z_{d-3,1}(j_1,i_1,j_2,i_2,j_3,i_3,r_3) \times \ten G_1(j_1, i_1, r_1)
\label{eq:ct11}\\
&\hat{\ten G_3}(r_2, j_3, i_3, r_3) 
= \sum_{r_1, j_2, i_2} \ten Z_{d-3,2}(r_1, j_2, i_2, j_3, i_3, r_3) 
\times \ten G_2(r_1, j_2, i_2, r_2)\label{eq:ct12}\\
&\dots \nonumber
\end{align}    
Note that the $\ten Z_{1,1}$ is shared to get $\hat{\ten G_1}$ and $\hat{\ten G_2}$, and $\ten Z_{2,1}$ is shared to get $\hat{\ten G_1}$ and $\hat{\ten G_3}$. 

\subsection{Use of PE}
To compute the gradient with respect to the tensor factors, as in equation (\ref{eq:ct8})-(\ref{eq:ct12}), 
we can reuse PE1 and PE2. In either case we can reshape the tensor in order to apply Equation \eqref{eq:pe1 contraction} or Equation \eqref{eq:pe2 contraction}.
If tensor contraction is executed along the the last dimension as in (\ref{eq:ct1})(\ref{eq:ct6})(\ref{eq:ct8})(\ref{eq:ct13})(\ref{eq:ct9}), PE1 (Equation \eqref{eq:pe1 contraction}) is used; otherwise, as in (\ref{eq:ct2})(\ref{eq:ct3})(\ref{eq:ct4})(\ref{eq:ct5})(\ref{eq:ct10})(\ref{eq:ct11})(\ref{eq:ct12}), PE2 (Equation \eqref{eq:pe2 contraction}) is used.
The way to shape the tensors is given in Table \ref{tab:pe}. When necessary we add a dimension with size one and/or perform a reshape operation so that our computation fits the appropriate PE tensor contraction. 
\begin{table}[t!]
\centering
\begin{tabular}{cc|cccc}
    \hline
    Eq. & PE & $a$ & $b$ & $c$ & $d$ \\
    \hline
    (\ref{eq:ct1}) & PE1 & $I_1I_2\dots I_{d-1}$ & 1 & $I_d$ & $R_{d-1}J_d$ \\
    (\ref{eq:ct2}) & PE2 & $I_1I_2\dots I_{d-2}$  & $I_{d-1}R_{d-1}$ & $J_d$ & $R_{d-2}J_{d-1}$\\
    (\ref{eq:ct3}) & PE2 & 1 & $I_1R_1$ & $J_2\dots J_d$ & $J_1$ \\
    (\ref{eq:ct4}) & PE2 & 1 & $J_1$ & $J_2\dots J_d$ & $I_1R_1$ \\
    (\ref{eq:ct5}) & PE2 & $I_1$ & $R_1J_2$ & $J_3\dots J_d$ & $I_2R_2$ \\
    (\ref{eq:ct6}) & PE1 & $I_1I_2\dots I_{d-1}$ & $R_{d-1}$ & $J_d$ & $I_d$ \\
    (\ref{eq:ct8}) & PE1 & $J_1 I_1\dots J_{d-1}I_{d-1}$ & 1 & $J_dI_d$ & $R_{d-1}$ \\
    (\ref{eq:ct13}) & PE1 & $J_1 I_1\dots J_{d-2}I_{d-2}$ & 1 & $J_{d-1}I_{d-1}$ & $R_{d-2}$\\
    (\ref{eq:ct9}) & PE1 & $J_1I_1$ & 1 & $J_2I_2R_2$ & $R_1$ \\
    (\ref{eq:ct10}) & PE2 & 1 & $J_1I_1$ & $J_2I_2R_2$ & $R_1$ \\
    (\ref{eq:ct11}) & PE2 & 1 & $J_1I_1$ & $J_2I_2J_3I_3R_3$ & $R_1$ \\
    (\ref{eq:ct12}) & PE2 & 1 & $R_1J_2I_2$ & $J_3I_3R_3$ & $R_2$ \\
    \hline
\end{tabular}
\caption{PE and operand of each expression.}
\label{tab:pe}
\end{table} 
\section{Detail about numerical experiment}
\label{sec:exp}
We used C++ to implement fixed point tensor contraction and used Pytorch to implement high level methods (ADAM, rank parameters update). In order to fit the requirement on the shape of tensors, we zero pad the input to $28 \times 32$ and decompose it to $7\times 4 \times 2 \times 16$. There are 512 neurons in the hidden layer decomposed into $4\times 4 \times 2 \times 16$ for the first layer, and $32 \times 16$ for the second layer. The output is decomposed into $1\times 16$. 
We trained this model for FashionMNIST dataset \citep{xiao2017/online}, which has the same shape and size as MNIST dataset but is more complicated and can better
represent modern machine learning tasks. To accelerate training, we pretrain the models on MNIST dataset. 
We train the model for 30 epochs and compare both standard floating-point computation in Pytorch (Floating) and our simulator (Fixed), and also compare the training methods with or without the low rank TT priors. We report the epoch with highest testing accuracy.
For run-time comparison, only the time on forward and backward propagation is included, as the rest part (optimizer) is the same on both devices.  

\end{document}